\RequirePackage{lineno}
\documentclass[12pt,a4paper]{article}

\usepackage{cancel}
\usepackage{cite}
\usepackage{graphicx}
\usepackage{epsfig}
\usepackage{amsmath,amsfonts,amssymb}
\usepackage{t1enc}

\newcommand{\ttbar}{t \bar t}
\newcommand{\uubar}{u \bar u}
\newcommand{\ddbar}{d \bar d}
\newcommand{\ttA}{t \bar t \gamma}
\newcommand{\afb}{A_\text{FB}}

\newcommand{\ac}{A_\text{C}}
\newcommand{\acA}{A_\text{C}^{t \bar t \gamma}}
\newcommand{\actt}{A_\text{C}^{t \bar t}}
\newcommand{\mttbar}{m_{t \bar t}}
\newcommand{\sla}[1]{/\!\!\!#1}

\begin{document}

\begin{center}
\begin{Large}
{\bf Shedding \textit{light} on the $\ttbar$ asymmetry: the photon handle}
\end{Large}

J. A. Aguilar--Saavedra$^{a}$, E. \'Alvarez$^{b}$, A. Juste$^{c,d}$, F. Rubbo$^{d}$ \\
\begin{small}
{\it $^a$ Departamento de F\'{\i}sica Te\'orica y del Cosmos, Universidad de Granada, E-18071 Granada, Spain} \\
{\it $^b$ CONICET, IFIBA Universidad de Buenos Aires, 1428 Buenos Aires, Argentina}\\
{\it $^c$ Instituci\'o Catalana de Recerca i Estudis Avan\c{c}ats (ICREA), E-08010 Barcelona, Spain} \\
{\it $^d$ Institut de F\'{\i}sica d'Altes Energies (IFAE), E-08193 Bellaterra, Barcelona, Spain}
\end{small}
\end{center}

\begin{abstract}
We investigate a charge asymmetry in $\ttA$ production at the LHC that provides complementary information to the measured asymmetries in $\ttbar$ production. We estimate the experimental uncertainty in its measurement at the LHC with 8 and 14 TeV. For new physics models that simultaneously reproduce the asymmetry excess in $\ttbar$ at the Tevatron and the SM-like asymmetry at the LHC, the measurement in $\ttA$ at the LHC could exhibit significant deviations with respect to the SM prediction.

\end{abstract}

\section{Introduction}

After several Tevatron legacy measurements with the full data set~\cite{Aaltonen:2012it,CDF:2013gna,Aaltonen:2013vaf,Abazov:2013wxa} and the most recent data from the Large Hadron Collider (LHC)~\cite{Chatrchyan:2011hk,Aad:2013cea,CMS:2013nfa}, the potential presence of new physics in $\ttbar$ production resulting in an anomalously large
forward-backward (FB) asymmetry remains intriguing. Measurements at Tevatron are above the Standard Model (SM) predictions~\cite{Ahrens:2011uf,Hollik:2011ps,Kuhn:2011ri,Bernreuther:2012sx}, for example $2.2$ standard deviations ($2.2\sigma$) in the case of the inclusive $t \bar t$ asymmetry $\afb$~\cite{Aaltonen:2012it}, which raises to $2.5\sigma$ for high $\ttbar$ invariant masses $m_{\ttbar} > 450$ GeV. But, on the other hand, measurements of the $\ttbar$ charge asymmetry $\ac$ at the LHC exhibit a good consistency with the SM expectations. At first glance, this is a surprising fact, because $\afb$ and $\ac$ origin from the very same partonic asymmetries $A_u$ and $A_d$ in $u \bar u \to \ttbar$, $d \bar d \to \ttbar$, respectively~\cite{AguilarSaavedra:2012va}. Therefore, $\afb$ and $\ac$ are tightly correlated~\cite{AguilarSaavedra:2011hz} in the simple models~\cite{AguilarSaavedra:2011ug} that were first proposed to explain the $\afb$ measurement~\cite{Djouadi:2009nb,Jung:2009jz,Cheung:2009ch,Shu:2009xf,Nelson:2011us}. In these models, with a single extra particle, either $A_u$ or $A_d$ is zero, or they have the same sign. Likewise, it is also expected that (isospin-symmetric) missing SM contributions that might increase the prediction of $\afb$ to make it closer to the experimental measurement, would also enhance $\ac$ and worsen the agreement with experimental data~\cite{Aguilar-Saavedra:2013rza}. Therefore, although Tevatron and LHC measurements are not incompatible, they are in tension within the SM and its simple extensions.

On the other hand, it is possible to have an excess at the Tevatron and no excess at the LHC if there is a cancellation of some type. There are two known ways to achieve this:
\begin{enumerate}
\item A cancellation between $\uubar$ and $\ddbar$ contributions~\cite{AguilarSaavedra:2012va}, as implemented for example in $s$-channel colour octet models~\cite{Drobnak:2012cz}. If the asymmetries $A_u$, $A_d$ have different sign, the higher importance of $\ddbar$ with respect to $\uubar$ at the LHC ---the ratio of cross sections $\sigma(\ddbar) / \sigma(\uubar)$ is a factor of three higher at the LHC than at the Tevatron--- makes it possible to have a positive asymmetry $\afb$, dominated by $A_u$, at the Tevatron, and a very small asymmetry $\ac$ at the LHC. 
\item A cancellation between $q \bar q \to t \bar t$  (with $q=u,d$) and $qg \to t \bar t j$ contributions, as for example in $Z'$ models~\cite{Leskow:2013wpa,Alvarez:2012ca,Drobnak:2012rb}. The $gq$ processes are irrelevant at the Tevatron but not at the LHC. There, a negative asymmetry in $ug \to t \bar t j$ could compensate a positive asymmetry in $u \bar u \to t \bar t$, if one considers not only $t \bar t$ production but also including additional jets. This model is disfavoured by the measurement of the high-$m_{t \bar t}$ tail at the LHC and may eventually be excluded with more precise measurements, but serves as a good benchmark for our discussion.
\end{enumerate}
Thus, the small SM-like asymmetry observed at the LHC may in principle result from two asymmetries of opposite sign (the naturalness of this mechanism is a different issue). But the addition of a final state photon completely changes this. An asymmetry can be defined in $\ttA$ production
\begin{equation}
\acA = \frac{N(\Delta |y|>0) - N(\Delta |y|<0)}{N(\Delta |y|>0) + N(\Delta |y|<0)} \,,
\label{ec:ac}
\end{equation}
with $\Delta |y| = |y_t| - |y_{\bar t}|$ the difference between the moduli of the rapidities of the top and antitop. That is, the same definition for the charge asymmetry as in $\ttbar$ production. In the SM, this asymmetry arises already at the tree level in $q \bar q \to \ttA$ due to the interference between diagrams where the photon is emitted from initial quarks and diagrams where it is emitted from final state quarks. At a centre-of-mass (CM) energy $\sqrt s = 8$ TeV the tree-level value is $\acA = -0.058$, and at $\sqrt s = 14$ TeV it is $\acA = -0.038$. Comparing to $\ttbar$ alone, the effect of the extra photon is twofold. First, it increases the $q \bar q$ fraction of total events $F_u+F_d$, $F_u \equiv \sigma(u \bar u)/\sigma$, $F_d \equiv \sigma(d \bar d)/\sigma$, thus reducing the washout due to symmetric $gg$ fusion. (For simplicity we ignore here $s \bar s$ and $c \bar c$ initial states, which are irrelevant for the discussion.) Second, and even more importantly, if there is a cancellation between new physics contributions to turn $\ac$ small, the addition of a photon is likely to break it in $\acA$. This is expected to be a general fact, and indeed is confirmed in this paper for the two aforementioned mechanisms.

\begin{figure}[htb]
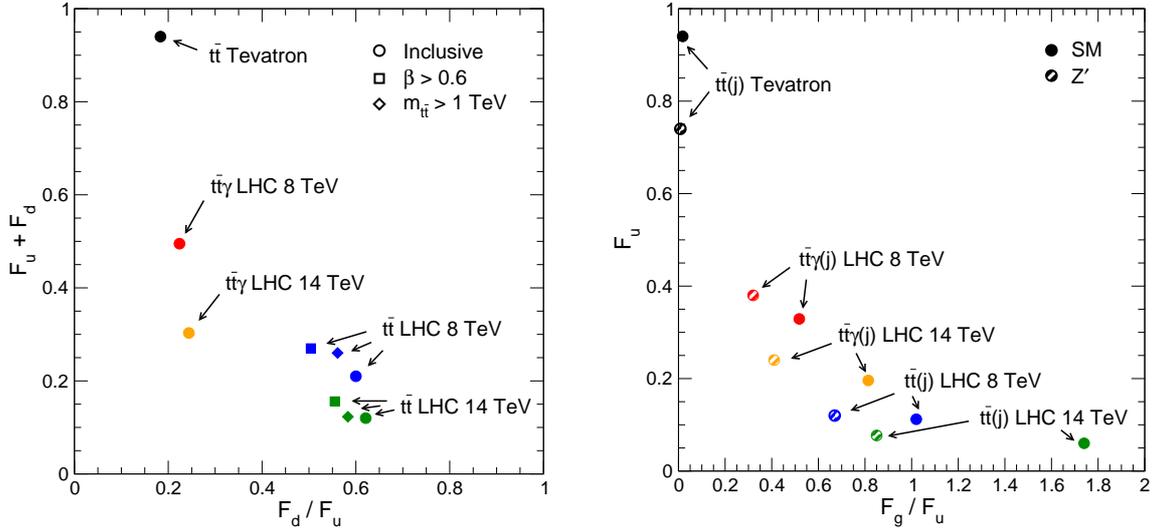

\begin{center}
\begin{tabular}{ccc}
\includegraphics[height=7cm,clip=]{Figs/r_ud.eps} & \quad
\includegraphics[height=7cm,clip=]{Figs/r_gu.eps} & \quad
\end{tabular}
\end{center}
\caption{Left: $F_d/F_u = \sigma(\ddbar) / \sigma(\uubar)$ and $q\bar q$ fraction $F_u + F_d$ for $\ttbar$ and $\ttA$ production in the SM. Right: $F_g/F_u = \sigma(ug)/\sigma(u\bar u)$ and $u \bar u$ fraction $F_u$ in  $t \bar t (j)$ and $t \bar t \gamma (j)$, in the SM and for the production and decay of a $Z'$.}
\label{fig:pdf}
\end{figure}

In the first case, namely a cancellation between $u\bar u$ and $d\bar d$ contributions at the LHC, such cancellation is broken mainly because the photon couples differently to up and down quarks, and $\uubar \to \ttA$ is enhanced with respect to  $\ddbar \to \ttA$. The effect can be clearly seen in Fig.~\ref{fig:pdf} (left), where we plot the $q\bar q$ fraction $F_u+F_d$ against the ratio $F_d/F_u$ for the SM processes. Kinematical cuts on the $\ttbar$ velocity $\beta$~\cite{oai:arXiv.org:1109.3710} and the $\ttbar$ invariant mass $m_{t\bar t}$ lead to small shifts in
$F_d/F_u$ and $F_u+F_d$ towards the Tevatron point. But, clearly, requiring a final state photon is much more effective: it renders $F_d/F_u$ very close to the Tevatron value (thus breaking a possible cancellation) and also increases the $q \bar q$ fraction by a factor of two, leading to larger asymmetries. For new physics contributions ---such as an $s$-channel colour octet--- that do not significantly alter the kinematics of $\ttbar$ and $\ttA$ production, the behaviour is the same. Note also that $F_d/F_u$ is nearly the same at the LHC with $\sqrt s =8$ and 14 TeV (as well as with 7 TeV), hence the possible cancellation will also operate at 14 TeV and the measurement of $\ac$ at this energy is expected to show good consistency with the SM. 

In the second case, the cancellation is broken due to two effects. First, there is a decrease in $F_g / F_u = \sigma(ug)/\sigma(u\bar u)$ in $t \bar t \gamma (j)$ with respect to $t \bar t(j)$ merely due to kinematics, which already happens in the SM. Secondly, there is a further decrease in the new physics $Z'$ contribution because in $u\bar u \to t \bar t$ it is more likely to emit a photon than in $ug \to t Z'$, since in the former there are more charged particles. We can observe this clearly in Fig.~\ref{fig:pdf} (right), where the solid points correspond to the SM, and the hashed points are the ones for the $Z'$ model,\footnote{For this model we focus on inclusive asymmetries and do not investigate the effect of kinematical cuts. It is well known for example that $Z'$ exchange enhances the high-mass $m_{t \bar t}$ tail at the LHC~\cite{oai:arXiv.org:1103.2765,AguilarSaavedra:2011ug}, and kinematical differences are important already in the $t \bar t$ differential distributions.} as described in Sect.~\ref{sec:3}. Notice also that $F_g/F_u$ is larger at 14 TeV, increasing the negative component of $\ac$, but the washout due to $gg$ fusion increases too, and generically one does not expect to have visible asymmetries in $t \bar t$ production at 14 TeV if a cancellation of this type takes place. Therefore,

As the main consequence of this simplified analysis, one can say that the requirement of an extra photon in $t \bar t$ production at the LHC ``reweights'' the new physics asymmetries in the different subprocesses, if present, recovering to a large extent the situation at the Tevatron. This is expected to be a rather general feature, and the exact numerical calculations carried out in this paper for two benchmark models confirm this semi-quantitative argument.

The remainder of this paper is organised as follows.  In Sect.~\ref{sec:2} we present the simulation details and event selection to suppress radiative top decays, and we estimate the expected uncertainty in $\acA$.  In Sect.~\ref{sec:3} we study $\acA$ for two different new physics models that explain the excess in $\afb$ and the absence of an excess in $\ac$.  Our conclusions and final remarks are in Sect.~\ref{sec:4}.

\section{Simulation details}
\label{sec:2}

In order to estimate the experimental uncertainty in the measurement of $\acA$, we focus on the semileptonic $t\bar{t}$ decay channel,
which constitutes the golden mode for this measurement owing to the relatively large branching ratio ($\sim 30\%$), manageable background
after $b$ tagging requirements, and the ability to fully reconstruct the $t\bar{t}$ kinematics. A complication arises from the fact that any realistic estimate 
should in principle take into consideration the full $2 \to 7$ calculation for the process $pp \to \ell \nu b q \bar{q}' \bar{b}\gamma+X$, since only a fraction of
the cross section for this final state originates from photon radiation off the initial state quarks or the top/antitop quarks (referred to as ``radiative top production''), that is, $pp \to t\bar{t}\gamma+X$ with subsequent decay of the $\ttbar$ pair. As we will see in the following, the cross section is actually dominated by photon radiation off the top decay 
products ($b$ quark, $W$ boson and its charged decay products), referred to here as ``radiative top decay''.  Therefore, it is necessary
to design an event selection capable of effectively suppressing the contribution from radiative top decays, which would dilute the sensitivity of $\acA$ to new physics in the production process.

We use {\sc MadGraph 5}~\cite{mgme5} to generate SM samples for the $2 \to 3$ process $pp \to t\bar{t}\gamma+X$, as well as the following two
$2 \to 7$ processes:  $pp \to t\bar{t}\gamma+X$ with $t\bar{t} \to \ell \nu b q \bar{q}' \bar{b}$ and $\ell=e,\mu$ (i.e. restricted to 
radiative top production), and $pp \to \ell \nu b q \bar{q}' \bar{b}\gamma+X$ (i.e. including both radiative top production and decay). In the latter case
we consider only resonant Feynman diagrams with two on-shell top quarks and two on-shell $W$ bosons.  The samples are generated for $\sqrt{s}=8$~TeV and 14 TeV using the CTEQ6L1 parton distribution
set~\cite{cteq6l1} and the event-by-event renormalisation and factorisation scale as implemented by default in  {\sc MadGraph 5}. A top quark mass of
$m_t=173$~GeV is assumed throughout this study. All samples are generated requiring for the photon transverse momentum $p_{\rm T}(\gamma)>20$~GeV and
pseudo-rapidity $|\eta(\gamma)|<2.5$. In the case of the $2 \to 7$ samples, additional requirements are made on final-state partons:
\begin{itemize}
\item $p_{\rm T}(\ell)>10$~GeV, $p_{\rm T}(j)>20$~GeV (in the following we will use $j$ to denote $q$ or $b$),
\item $|\eta(\ell)|<2.5$,  $|\eta(j)|<5.0$,
\item lego-plot distance $\Delta R(\ell,\gamma)>0.4$, $\text{min}\{\Delta R(\ell,j)\}>0.4$, $\text{min}\{\Delta R(\gamma,j)\}>0.4$, $\text{min}\{\Delta R(j,j)\}>0.4$.
\end{itemize}
We perform our studies at the parton level but attempt to obtain reasonable estimates for the event selection efficiency and dilution in the measurement of $\acA$. 
We do not include backgrounds, although experimental studies of $\ttA$ production at the LHC~\cite{ATLASttA, CMSttA} find that the overall background 
can be comparable to the signal, and dominated by $t\bar{t}$ production with an electron or jet being identified as a photon. Although such background 
contribution is large, its associated asymmetry can be precisely measured in dedicated data control samples. Estimating the effect of the background
on the statistical and systematic uncertainty on $\acA$ is beyond the scope of this exploratory work and should be done in the context of the actual
experimental analyses.

Events are required to fulfill basic preselection requirements aimed at selecting the semileptonic signature:
\begin{itemize}
\item exactly one lepton (electron or muon) with transverse momentum $p_{\rm T}>20$~GeV and pseudorapidity $|\eta|<2.5$;
\item missing transverse momentum $\sla{p}_{\rm T}>20$~GeV;
\item four quarks with $p_{\rm T}>25$~GeV and $|\eta|<4.5$;
\item exactly one photon with $p_{\rm T}>20$~GeV and $|\eta|<2.5$;
\item $\text{min}\{\Delta R(\ell,j)\}>0.4$, $\text{min}\{\Delta R(\ell,\gamma)\}>0.4$, $\text{min}\{\Delta R(j,j)\}>0.4$.
\end{itemize}
With these requirements, the leading-order (LO) effective cross section at $\sqrt{s}=14$~TeV is 0.13~pb in the case of the $2 \to 7$ calculation without
radiative top decays, while it is 0.23 pb for the full $2 \to 7$ calculation, clearly showing the need to have additional cuts to suppress the large contribution
from radiative top decays. This can be accomplished by a combination of cuts: 
\begin{itemize}
\item tighter $\Delta R$ requirements between photon and charged top quark decay products: 
$\Delta R(\gamma, \ell)>1.0$, $\text{min}\{\Delta R(\gamma,q)\}>0.7$, $\text{min}\{\Delta R(\gamma,b)\}>0.5$;
\item veto radiative $W$ decays: $m(jj\gamma)>90$~GeV, $m_{\rm T}(\ell\gamma;\sla{p}_{\rm T})>90$~GeV,
where $m(jj\gamma)$ is the invariant mass of the $jj\gamma$ system and $m_{\rm T}(\ell\gamma;\sla{p}_{\rm T})$ is the cluster transverse mass defined as
$$
m_{\rm T}^2(\ell\gamma;{p}_{\rm T})=\left(\sqrt{p_{\rm T}^2(\ell\gamma)+m^2(\ell\gamma)}+ \sla{p}_{\rm T}\right)^2-\left(\vec{p}_{\rm T}(\ell\gamma)+\vec{\sla{p}}_{\rm T}\right)^2 \,,
$$
with analogous definitions for particles other than the photon and the charged lepton;
\item veto radiative top decays: reject events satisfying either of the following conditions:
\begin{enumerate}
\item $m_{\rm T}(b_{1,2}\ell\gamma;\sla{p}_{\rm T}) < m_t + 20~{\rm GeV}$ and $m_t - 20~{\rm GeV} < m(b_{2,1}jj) < m_t + 20~{\rm GeV}$;
\item
$m_{\rm T}(b_{1,2}\ell;\sla{p}_{\rm T}) < m_t + 20~{\rm GeV}$ and $m_t - 20~{\rm GeV} < m(b_{2,1}jj\gamma) < m_t + 20~{\rm GeV}$,
\end{enumerate}
where $b_1, b_2 = b, \bar{b}$, and $b_1 \neq b_2$;
\item consistency with radiative top production:
$m_{\rm T}(b_{1,2}\ell;\sla{p}_{\rm T}) < m_t + 20~{\rm GeV}$ and $m_t - 20~{\rm GeV} < m(b_{2,1}jj) < m_t + 20~{\rm GeV}$.
\end{itemize}

\begin{figure}[htb]
\begin{center}
\begin{tabular}{ccc}
\includegraphics[height=5.cm,clip=]{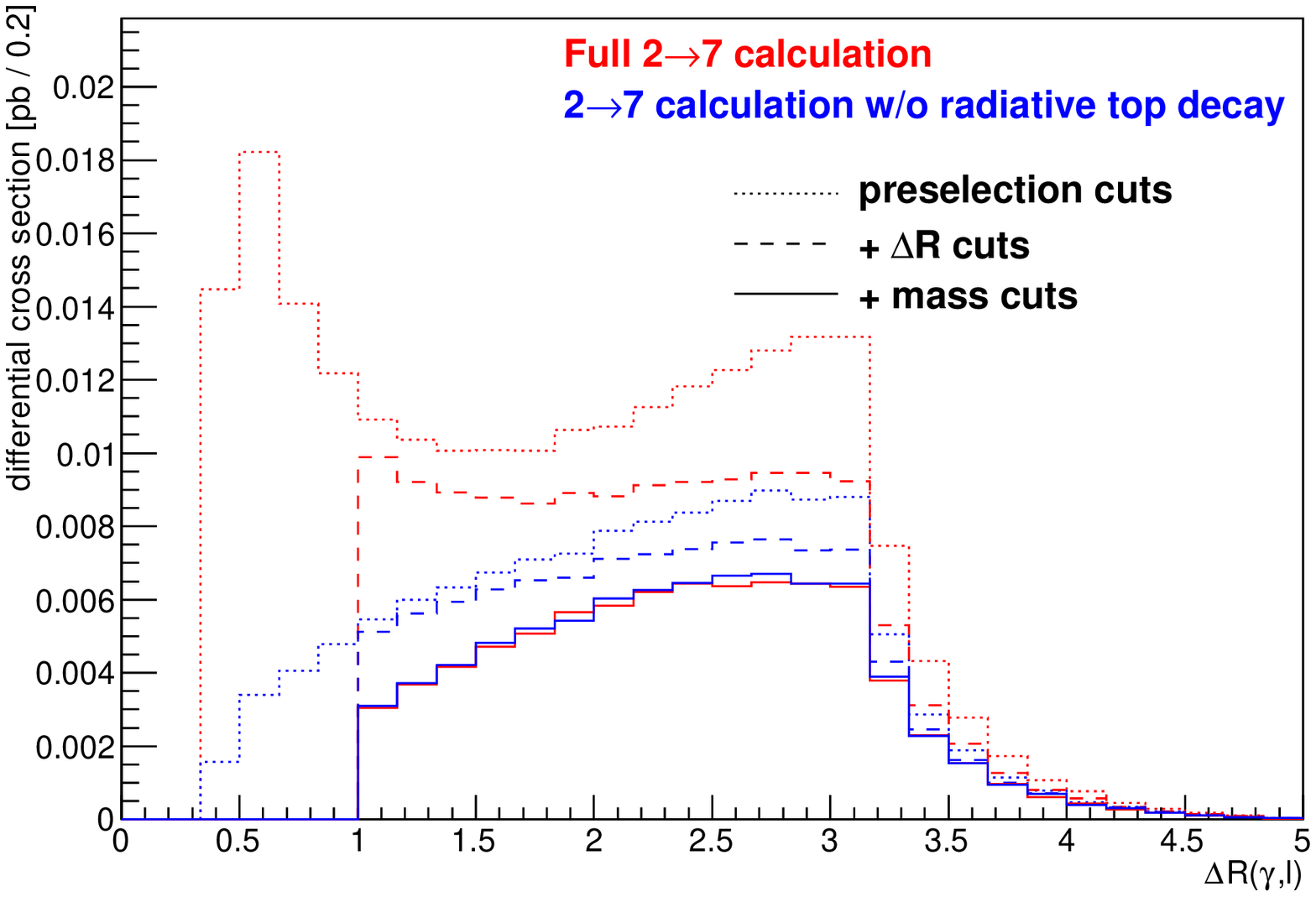} & \quad
\includegraphics[height=5.cm,clip=]{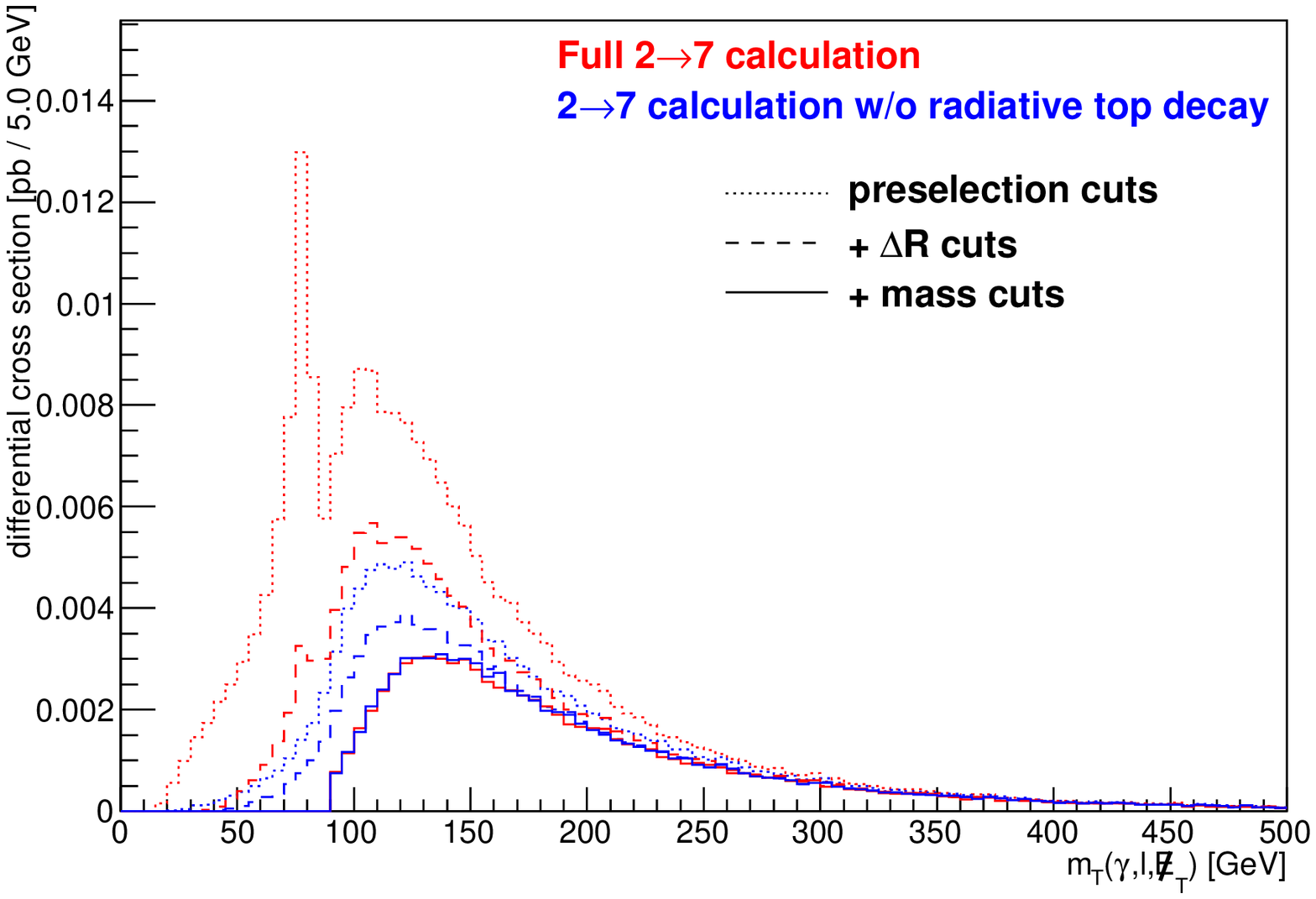} \\[2mm]
\includegraphics[height=5.cm,clip=]{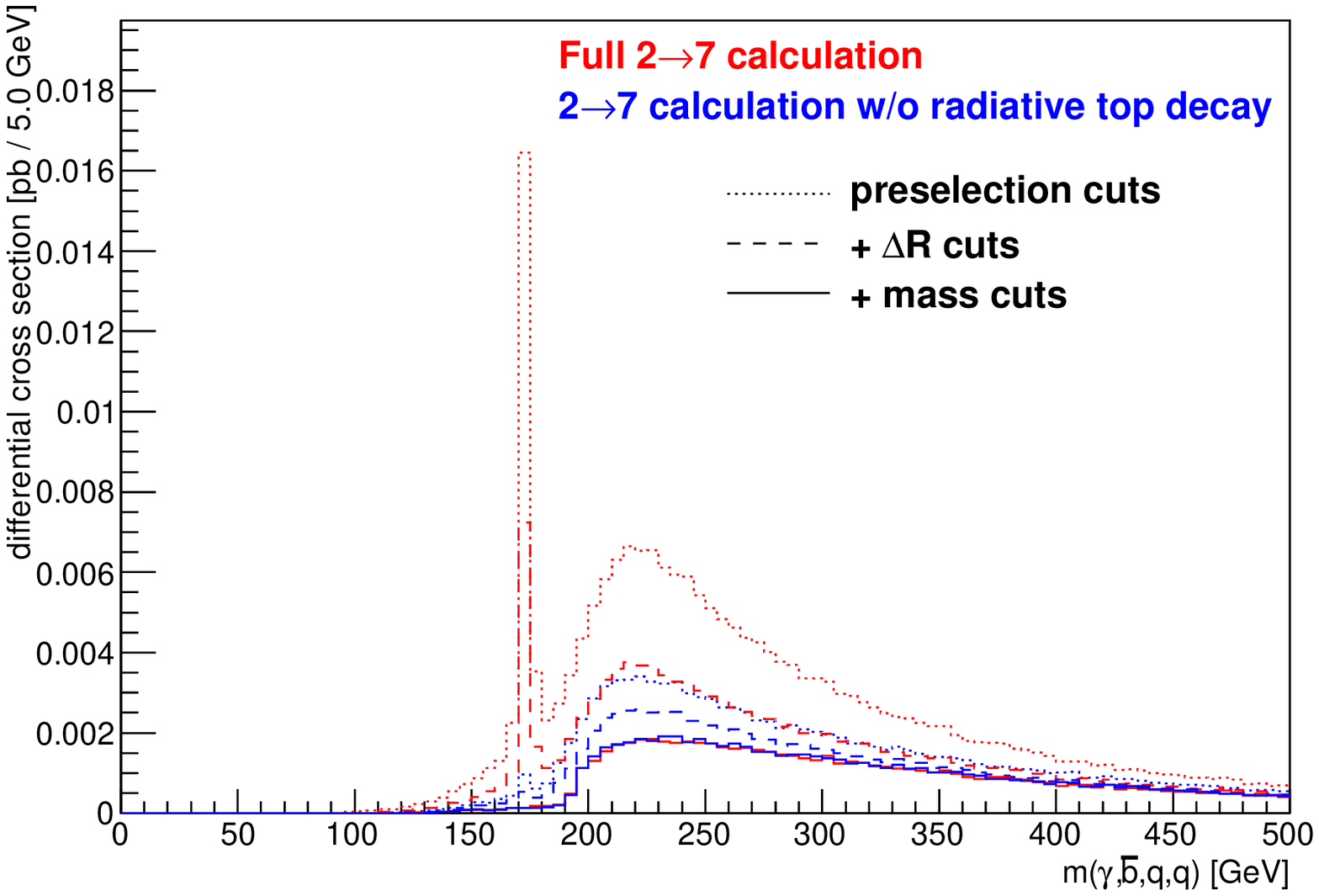} & \quad
\includegraphics[height=5.cm,clip=]{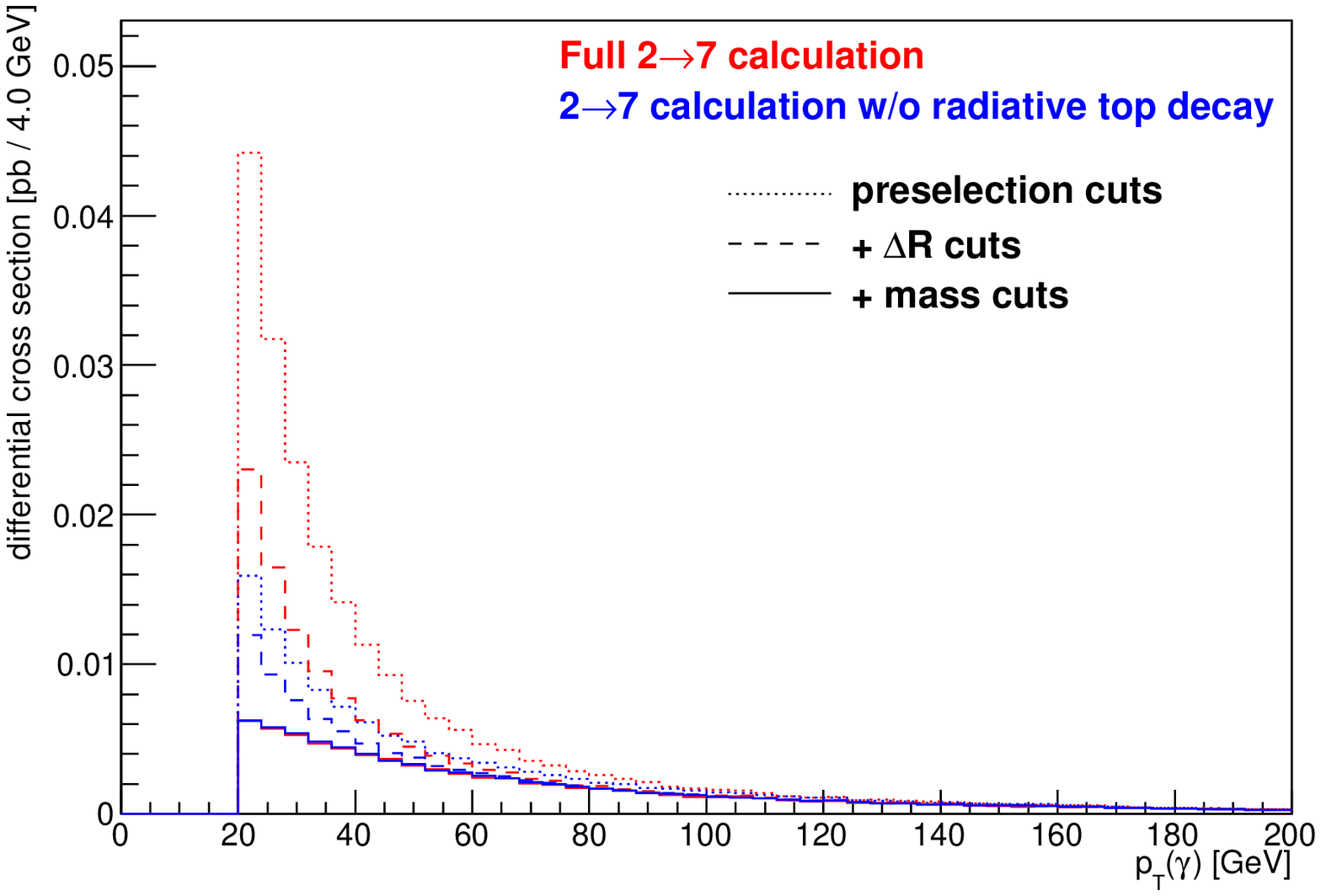} \\
\end{tabular}
\end{center}
\caption{Comparison of the differential cross sections between the
full $2 \to 7$ calculation (red) and the $2 \to 7$ calculation without radiative top decays (blue)
as a function of several kinematic variables. The different line styles
correspond to different stages of the event selection (see text for details): preselection cuts (dotted), preselection+$\Delta R$ cuts (dashed),
all selection cuts (solid).
}
\label{fig:distrib}
\end{figure}

\begin{figure}[htb]
\begin{center}
\begin{tabular}{ccc}
\includegraphics[height=5.cm,clip=]{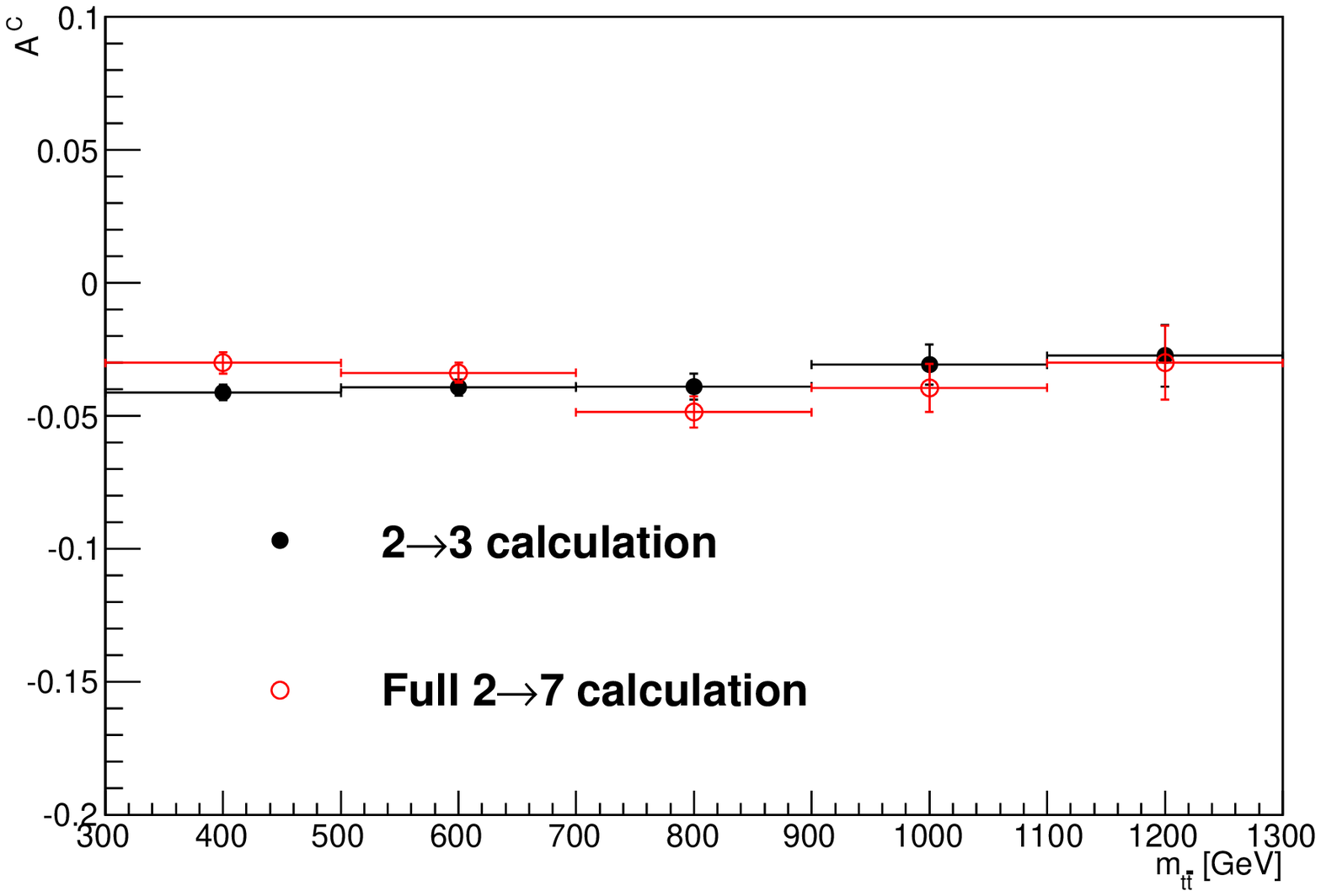} & \quad
\includegraphics[height=5.cm,clip=]{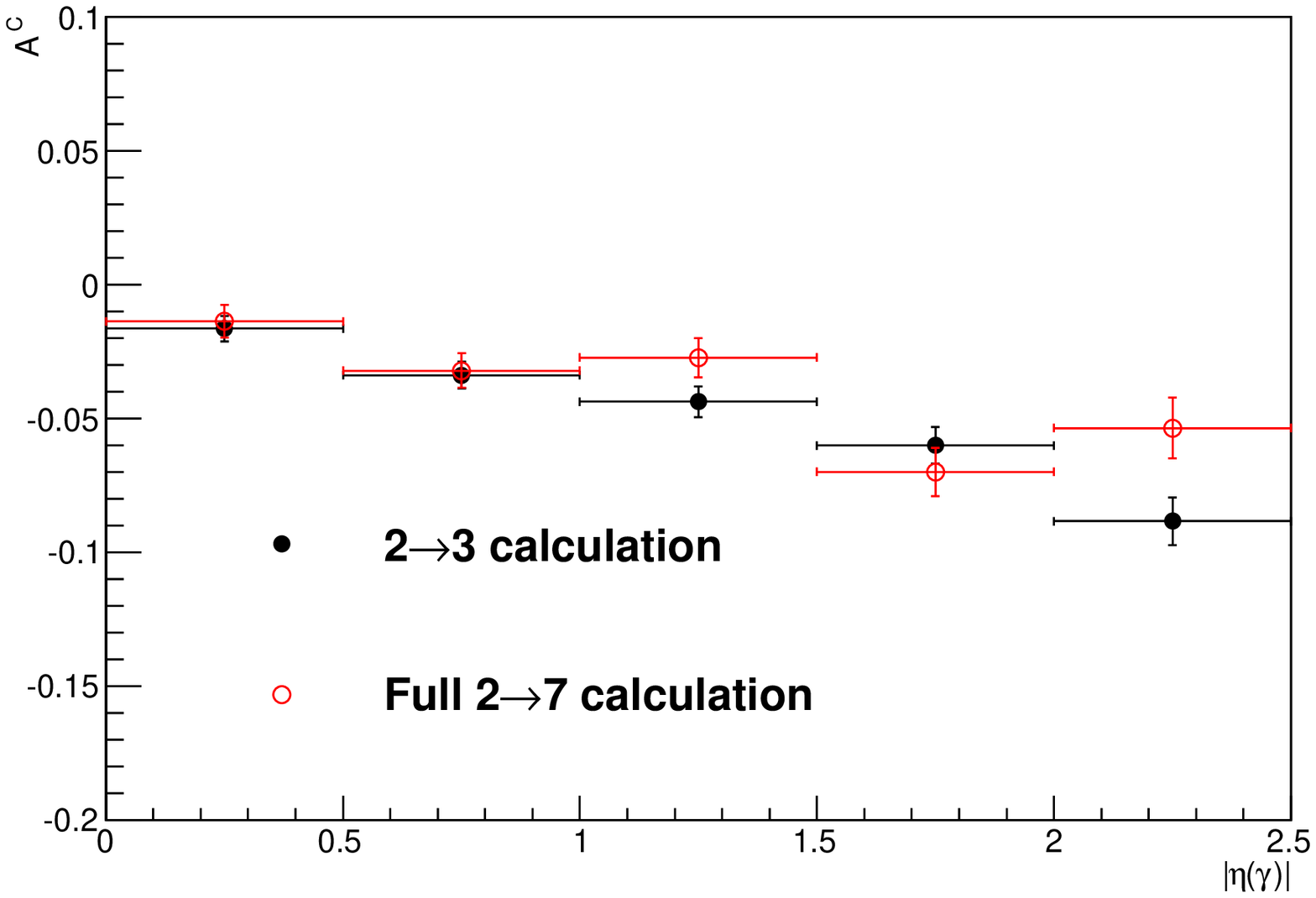} 
\end{tabular}
\end{center}
\caption{Left: $\acA$ as a function of $\mttbar$ at $\sqrt{s}=14$~TeV. Right: $\acA$ as a function of $|\eta(\gamma)|$  at $\sqrt{s}=14$~TeV.
The predicted $\acA$ is compared between the $2 \to 3$ calculation with just the photon cuts (solid black points), and
the full $2 \to 7$ calculation including all selection cuts (open red points). The error bars reflect the uncertainty from Monte Carlo statistics.
See text for details on the exact selection requirements.}
\label{fig:asym}
\end{figure}

After all requirements, the effective cross section at $\sqrt s = 14$ TeV becomes 0.083 pb in the case of the $2 \to 7$ calculation without
radiative top decays, while it is also 0.083 pb for the full $2 \to 7$ calculation, demonstrating that the contribution from radiative top decays
has been effectively eliminated. The effect of the sequential application of these cuts in some of the relevant kinematic distribution can 
be visualised in Fig.~\ref{fig:distrib}, where it is shown that after all requirements there is good agreement in both normalisation and
shape between the full $2 \to 7$ calculation and the $2 \to 7$ calculation without radiative top decays.
As a final check, we have compared the $\acA$ between the $2 \to 3$ calculation with just the photon cuts, and
the full $2 \to 7$ calculation including all selection cuts. The inclusive asymmetries are $\acA = -0.039(2)$ and $\acA = -0.035(2)$, respectively, where the
numbers in parentheses represent the uncertainty for the limited Monte Carlo statistics. Reasonable agreement is also found for
$\acA$ differentially as a function of $\mttbar$ and $|\eta(\gamma)|$, as shown in Fig.~\ref{fig:asym}.

Therefore, in the remainder of this paper we will only consider the simpler $2 \to 3$ calculation
for the $\ttA$ process together with the selection efficiency estimated from the above study and summarised below.
For a more realistic estimate, the LO cross section for the  $2 \to 3$ calculation is multiplied by a k-factor of 1.75,
in order to reproduce the NLO cross section of Ref.~\cite{Melnikov}. Therefore, the total $\ttA$ cross section for $\sqrt{s}=8$~TeV is 0.68 pb, and for $\sqrt{s}=14$~TeV it is 2.73 pb, 
corresponding to the requirements of $p_{\rm T}(\gamma)>20$~GeV and $|\eta(\gamma)|<2.5$.  

The semileptonic branching ratio is taken to be 30\%.
The total selection efficiency for semileptonic decays, including the effect of the preselection cuts as well the additional cuts to veto radiative top decay,
is estimated to be $\sim 20\%$ using a sample for the $2 \to 7$ process without radiative top decays, including only generator-level cuts on the
photon of $p_{\rm T}(\gamma)>20$~GeV and $|\eta(\gamma)|<2.5$. 
In order to obtain a more realistic estimate,  we also assume a total efficiency for lepton triggering and identification of 70\%, 
a photon identification efficiency of 85\% and a per-jet $b$ tagging efficiency of 70\%, resulting in an efficiency for a requirement of at least one $b$-tagged jet
of 90\%. The product of the above numbers results in a total efficiency times branching ratio of 3.2\%. Therefore, the expected number of selected
events for a total integrated luminosity of 20 fb$^{-1}$ at $\sqrt s = 8$ TeV is $\sim 440$, resulting in an expected statistical uncertainty on $\acA$ of $\pm 0.045$. At a CM energy of $\sqrt{s}=14$~TeV with 100 fb$^{-1}$ the expected number of events is $\sim 8800$, giving a statistical uncertainty on $\acA$ of $\pm 0.01$.

In the next section we will compute $\acA$ (see Eq.~\ref{ec:ac}) from $y_t$  and $y_{\bar t}$ at the parton level using the $2 \to 3$ calculation for the $pp \to t\bar{t}\gamma+X$ process in the presence of new physics.
Experimentally, the charge asymmetry would be computed after a kinematic reconstruction of the $t\bar{t}$ system in $\ttA$ events similar to that used for the 
measurement of $\actt$ in $t\bar{t}$ production~\cite{Aad:2013cea,CMS:2013nfa}. In the case of experimental measurements of $\actt$, the fraction of events with 
the sign of $\Delta  |y|$ correctly reconstructed is $\sim 75\%$. Using the simplified $t\bar{t}$ event reconstruction of Ref.~\cite{oai:arXiv.org:1109.3710}, 
we have verified that the fraction of events with the sign of $\Delta |y|$ correctly reconstructed is
in fact $\sim 5\%$ higher in absolute term for $\ttA$ compared to $t\bar{t}$ production, possibly related to the higher longitudinal boost of the $t\bar{t}$ system in the case of $\ttA$ 
events from the higher fraction of $q\bar{q}$-initiated production. Nevertheless, we will conservatively assume the same fraction of 75\%, which corresponds to a 
``dilution'' factor $D=2 \times 0.75 - 1 = 0.5$. Such dilution results in a reduction by a factor of two of the measured $\acA$ which is effectively corrected for by the unfolding procedure used 
in the experimental analyses, and can be taken into account here by an increase by a factor of two of the statistical uncertainty. Then, for $\sqrt{s}=8$~TeV and an integrated luminosity of 20 fb$^{-1}$, the resulting statistical uncertainty is $\pm 0.09$, and at $\sqrt{s}=14$~TeV with integrated luminosities of 100 fb$^{-1}$ (400 fb$^{-1}$) it is $\pm 0.02$ ($\pm 0.01$). Our estimates do not 
include systematic uncertainties since, based on the present experience with the $\ac$ measurements, those are expected to be small ($\leq 0.005$). In any case, a careful assessment of systematic uncertainties would require a detailed analysis involving experimental simulations that are beyond the scope of this work.

\section{Asymmetries in $\ttA$ and $\ttbar$}
\label{sec:3}

The importance of $\acA$ to provide a complementary probe of asymmetric $\ttbar$ production at the Tevatron is shown here with two benchmark examples. They correspond to the two types of cancellations between new physics contributions that may yield a small $\ac$ at the LHC. For our Monte Carlo calculations of $\ttbar$ with a colour octet we use {\sc Protos}~\cite{AguilarSaavedra:2008gt}.
For $\ttA(j)$ production in the colour octet and with an extra $Z'$ we use {\sc MadGraph 5}~\cite{mgme5} coupled to {\sc Pythia}~\cite{pythia} in its deafult tune with a MLM matching scheme~\cite{oai:arXiv.org:0706.2569} implemented to avoid double counting in the extra jet. In both models the asymmetries $\acA$ are calculated at leading order. For the comparison with experimental data, especially at $\sqrt s = 14$ TeV, next-to-leading order (NLO) computations would be required, at least for the SM asymmetries. (The asymmetries including new physics contributions could be approximated, as usual, by summing the SM asymmetries at NLO and the new physics contributions at LO.) In any case, our leading-order calculations are sufficient to demonstrate the importance of this asymmetry and its potential to exhibit deviations from the SM predictions.

\subsection{$\acA$ for a colour octet}
\begin{figure}[htb]
\begin{center}
\begin{tabular}{ccc}
\includegraphics[height=5.2cm,clip=]{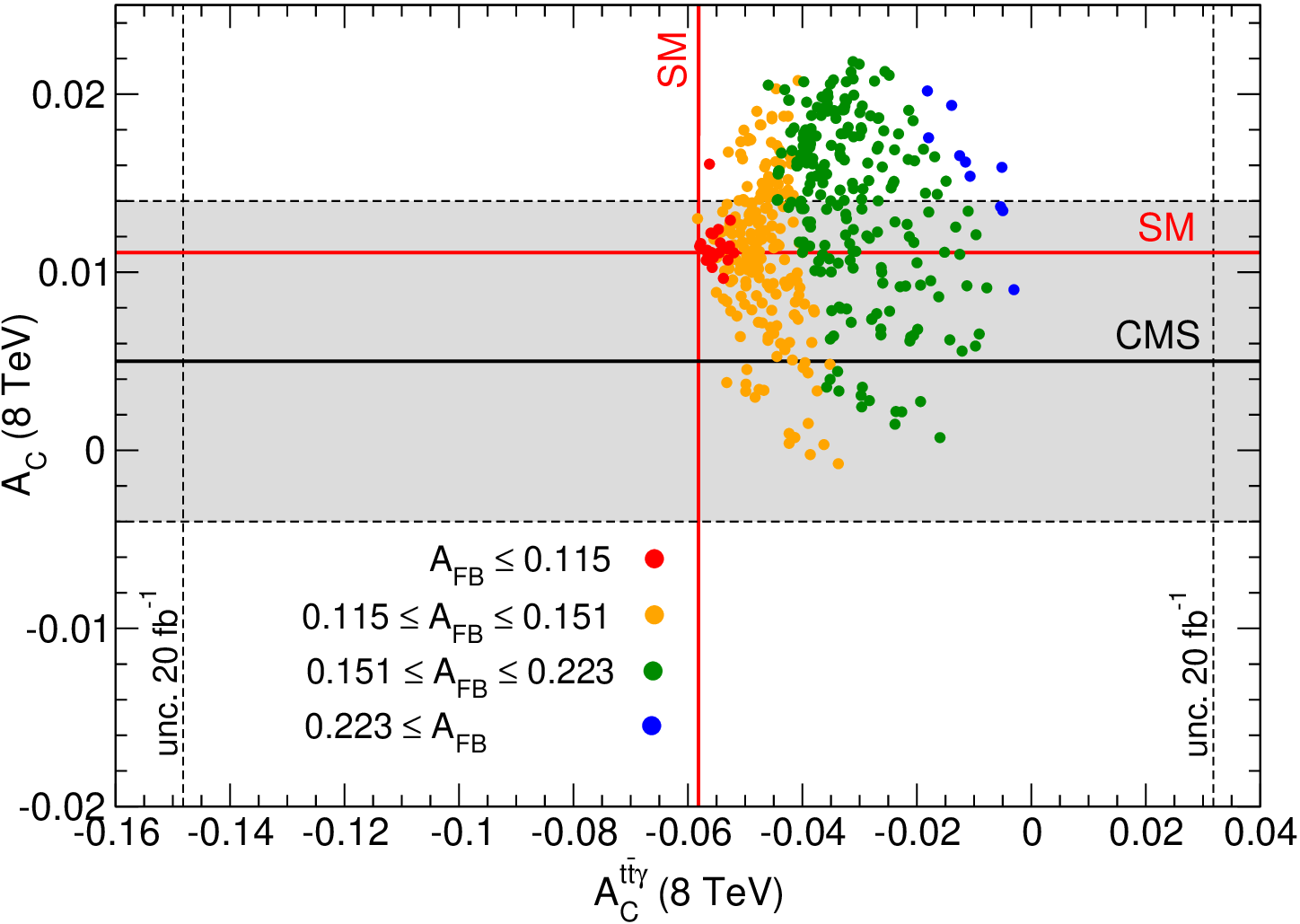} & \quad
\includegraphics[height=5.2cm,clip=]{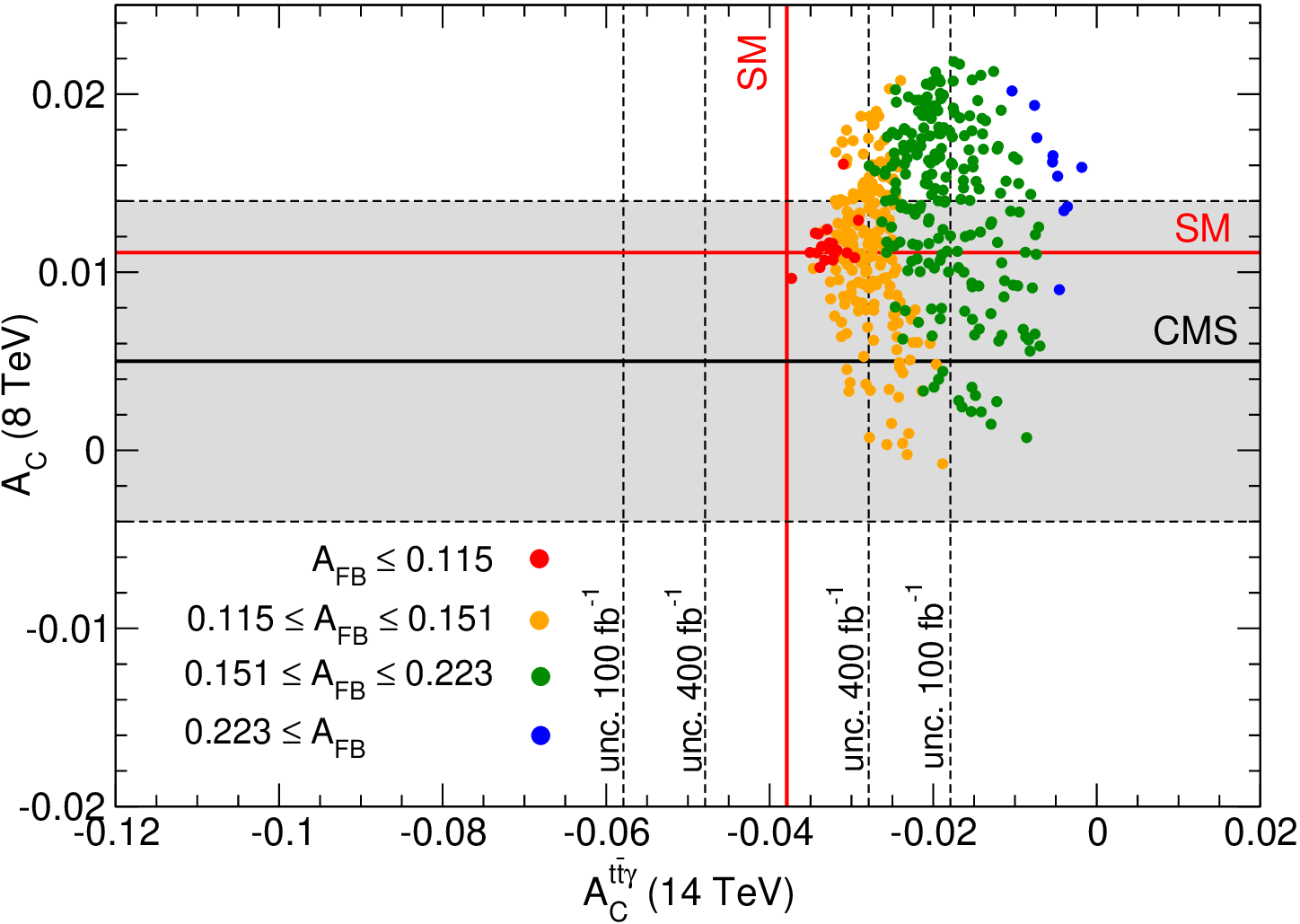} & \quad
\end{tabular}
\end{center}
\caption{Charge asymmetry $\acA$ (at 8 and 14 TeV) versus $\ac$ for a colour octet.}
\label{fig:ttAvstt}
\end{figure}

We scan the parameter space of couplings of a {\it light} colour octet~\cite{Barcelo:2011vk,Tavares:2011zg,Alvarez:2011hi,AguilarSaavedra:2011ci,Krnjaic:2011ub}, with a fixed mass $M=250$ GeV and a large width $\Gamma/M=0.2$, that is assumed in order to comply with dijet pair constraints~\cite{Gross:2012bz,Gresham:2012kv}. (This width may originate from additional decays to non-SM particles.)  This scan is a particular subset of a more general parameter space scan to be presented and discussed elsewhere~\cite{inprep}. For simplicity, here we take the vector couplings of the up and down quarks to zero. The parameter space points considered here have a global agreement with $\ttbar$ experimental data at the level of $1\sigma$ or better, including several observables such as cross sections and asymmetries.
The details of the fit are not essential here for our purposes, and it is sufficient to mention that the resulting points comprise high as well as low values of $\afb$, that might ---or not--- explain the Tevatron anomaly. Analogously, the resulting values of $\ac$ span a $\pm 2\sigma$ range above and below the SM predictions for 7 and 8 TeV. For these points, we present in Fig.~\ref{fig:ttAvstt} the asymmetry $\ac$ at 8 TeV versus $\acA$ at 8 TeV (left) and 14 TeV (right). The shaded horizontal area corresponds to the CMS measurement $\ac=0.005 \pm 0.009$~\cite{CMS:2013nfa} and its $1 \sigma$ uncertainty. The vertical lines represent the SM value of $\acA$ and its expected uncertainty, estimated in section~\ref{sec:2}. The points are coloured according to the value of the Tevatron asymmetry $\afb$. From left to right: red points have $\afb$ below $-2\sigma$ of the naive Tevatron average $A_\text{FB}^\text{exp} = 0.187 \pm 0.036$; orange points have $\afb$ between $-2 \sigma$ and $-1\sigma$; green points have $\afb$ between $-1\sigma$ and $1\sigma$; blue points have asymmetries above $2\sigma$. In all cases the SM contribution to the asymmetries~\cite{Bernreuther:2012sx} is summed to the ones resulting from new physics.

The difference found between 8 TeV and 14 TeV is given by the much better statistics at the latter CM energy, which by far compensates the slightly smaller $\acA$ due to the increased $gg$ cross section. (As pointed out before, the ratio $F_d/F_u$ is nearly the same at these two energies.) Interestingly, for the parameter space points (in green) that predict an asymmetry $\afb$ within $\pm 1 \sigma$ of the Tevatron average, the asymmetry in $\ttA$ departs  $1\sigma$ or more above the SM value. This happens irrespectively of the value of the asymmetry in $\ttbar$ production $\ac$, that may even have the SM value. We also note that for all points the increase in $\acA$ with respect to the SM value is positive. This is expected from the behaviour shown in Fig.~\ref{fig:pdf} (left), and is clearly because the increase in $\afb$ with respect to the SM value is also positive, as preferred by the fit to the Tevatron measurements.

\subsection{$\acA$ for a $Z'$}

We have chosen a reference point in the extra $Z'$ model that represents the region of parameter space compatible with the relevant observables as stated in the study in Ref.\cite{Alvarez:2013jqa}.  We have set $M_{Z'}=275$ GeV, $g_{utZ'}=0.7$ and an invisible $Z'$ branching ratio ${\mathcal B}(Z' \to \mbox{invisible})=3/4$. This model predicts a charge asymmetry at the LHC of $\ac = 0.015$ at 8 TeV, which is compatible with CMS measurement~\cite{CMS:2013nfa} and SM predictions.   However, when an extra photon is required, its prediction yields $\acA = -0.048$ and $\acA  = -0.018$ at 8 and 14 TeV, respectively, which again is an excess over the SM expected asymmetry (compare to SM prediction in Fig.~\ref{fig:ttAvstt}).  We see that also in this case, the 8 TeV measurement of $\acA$ would not differentiate this model from SM, but we can expect that 400 fb$^{-1}$ at 14 TeV could be able to do it.

\subsection{$\ac$ at 14 TeV}

Finally, in order to stress the importance of $\acA$, we have checked that in both models the prediction of $\ac$ in $\ttbar$ production at 14 TeV is always consistent with the SM, taking an experimental uncertainty of $\pm 0.005$. This is expected from the arguments given in the introduction. If a cancellation between $u \bar u$ and $d \bar d$ contributions takes place to render the LHC asymmetry small at 7 and 8 TeV, it should also operate at 14 TeV, where $F_d/F_u$ is similar. Or, if there exists a cancellation between $u \bar u$ and $gu$, at 14 TeV this cancellation is relaxed by the larger $F_g/F_u$ but the resulting asymmetry is small anyway.

\section{Conclusions}
\label{sec:4}

After few years of measurements with increasing precision, the excess asymmetry $\afb$ in $\ttbar$ production at the Tevatron and the SM-like charge asymmetry $\ac$ in $\ttbar$ production at the LHC continue to be puzzling. If new data confirm the current central values and reduce their uncertainties, it is unlikely that both can be explained within the SM (also with additional isospin-conserving contributions yet unaccounted for) and simple extensions. Barring the possibility of unknown systematic uncertainties, the only way to explain these measurements is by means of some cancellation of new physics contributions at the LHC, to give approximately the SM value.

A previous attempt to solve this puzzle has been the introduction of the collider-independent asymmetries~\cite{AguilarSaavedra:2012va,AguilarSaavedra:2012rx} $A_u$, $A_d$ that are the same at the Tevatron and the LHC, and whose measurement ---quite demanding from the experimental point of view--- could settle the issue. Also, the determination of ratios of differential asymmetries~\cite{Falkowski:2012cu,Carmona:2014gra} would be useful to uncover possible anomalies.
In this paper we have pointed out that, if the pattern of large $\afb$ and SM-like $\ac$ corresponds to some cancellation, such cancellation should not occur in $\ttA$ production at the LHC, where the additional photon alters the balance between new physics contributions with respect to $\ttbar$ alone. This fact turns extremely interesting the measurement of a charge asymmetry in $\ttA$ production. Moreover, even if the Tevatron excess is not due to new physics, the measurement of $\acA$ could bring light into the nature of the Tevatron anomaly.

We have studied the feasibility of this measurement and we have shown that the unwanted contributions to the final state of interest, where the photon is radiated off the top quark decay products, can be effectively suppressed with a judicious set of kinematical cuts. We have then estimated the experimental uncertainty on the measurement of $\acA$, to investigate the potential to observe deviations from the SM.

At 7 and 8 TeV, the statistics is likely not enough to reach a conclusive statement. At 14 TeV, we have shown that there are good chances of observing a deviation from the SM prediction in the case of a colour octet and in the extra $Z'$ model. In both cases we do find an excess in $\acA$ with respect to the SM ---and we argue that this should be in general expected for new physics models that yield an excess in $\afb$.  On the other hand, the predictions for the charge asymmetry $\ac$ in $\ttbar$ production at 14 TeV are close to the SM expectation. Therefore ---aside more demanding doubly-differential studies of the $\beta$ and $m_{t \bar t}$ dependence of $\ac$~\cite{AguilarSaavedra:2012va} or ratios of differential asymmetries~\cite{Falkowski:2012cu,Carmona:2014gra}--- the measurement of a charge asymmetry in $t \bar t \gamma$ seems the best avenue to explore FB-asymmetric $t \bar t$ production in the upcoming LHC run.

\section*{Acknowledgements}
This work has been supported by MICINN projects FPA2010-17915 and FPA2012-38713, by
Junta de Andaluc\'{\i}a projects FQM 101, FQM 03048 and FQM 6552, by FCT project EXPL/FIS-NUC/0460/2013, and by ANPCyT PICT 2011-0359.

\end{document}